\input harvmac
\input graphicx

\def\Title#1#2{\rightline{#1}\ifx\answ\bigans\nopagenumbers\pageno0\vskip1in
\else\pageno1\vskip.8in\fi \centerline{\titlefont #2}\vskip .5in}
%

%
%
\ifx\includegraphics\UnDeFiNeD\message{(NO graphicx.tex, FIGURES WILL BE IGNORED)}
\def\figin#1{\vskip2in}
\else\message{(FIGURES WILL BE INCLUDED)}\def\figin#1{#1}
\fi
\def\Fig#1{Fig.~\the\figno\xdef#1{Fig.~\the\figno}\global\advance\figno
 by1}
%
%
%
%
\def\Ifig#1#2#3#4{
\goodbreak\midinsert
\figin{\centerline{
\includegraphics[width=#4truein]{#3}}}
\narrower\narrower\noindent{\footnotefont
{\bf #1:}  #2\par}
\endinsert
}
%
%
\font\ticp=cmcsc10
\def\subsubsec#1{\noindent{\undertext { #1}}}
\def\undertext#1{$\underline{\smash{\hbox{#1}}}$}

\def\calh{{\cal H}}
\def\hbh{\calh_{bh}}
\def\hnear{\calh_{near}}
\def\hfar{\calh_{far}}
\def\hreg{\calh_{reg}}
\def\hcore{\calh_{core}}

\def\eg{{\it e.g.}}
\def\roughly#1{\mathrel{\raise.3ex\hbox{$#1$\kern-.75em\lower1ex\hbox{$\sim$}}}}

\def\svn{S_{\rm vN}}
\def\sbh{S_{\rm bh}}
\def\sBH{S_{\rm BH}}
\def\rads{R_{\rm AdS}}
\def\tsc{T_{\rm sc}}
\def\tx{T_{\rm xfer}}
\def\tpage{T_{\rm Page}}
\overfullrule=0pt
%
%

\lref\Hawkrad{
  S.~W.~Hawking,
  ``Particle Creation By Black Holes,''
  Commun.\ Math.\ Phys.\  {\bf 43}, 199 (1975)
  [Erratum-ibid.\  {\bf 46}, 206 (1976)].
}
\lref\BHMR{
  S.~B.~Giddings,
  ``Black holes and massive remnants,''
Phys.\ Rev.\  {\bf D46}, 1347-1352 (1992).
[hep-th/9203059].
}
\lref\thooholo{
  G.~'t Hooft,
  ``Dimensional reduction in quantum gravity,''
  arXiv:gr-qc/9310026.
}
\lref\sussholo{
  L.~Susskind,
  ``The World As A Hologram,''
  J.\ Math.\ Phys.\  {\bf 36}, 6377 (1995)
  [arXiv:hep-th/9409089].
}
\lref\StVa{
  A.~Strominger and C.~Vafa,
  ``Microscopic origin of the Bekenstein-Hawking entropy,''
Phys.\ Lett.\ B {\bf 379}, 99 (1996).
[hep-th/9601029].
}
\lref\NVNL{
  S.~B.~Giddings,
  ``Nonviolent nonlocality,''
[arXiv:1211.7070 [hep-th]].
}
\lref\AMPS{
  A.~Almheiri, D.~Marolf, J.~Polchinski and J.~Sully,
  ``Black Holes: Complementarity or Firewalls?,''
[arXiv:1207.3123 [hep-th]].
}
\lref\modelsone{
  S.~B.~Giddings,
  ``Models for unitary black hole disintegration,''
Phys.\ Rev.\ D {\bf 85}, 044038 (2012).
[arXiv:1108.2015 [hep-th]].
}
\lref\modelstwo{
  S.~B.~Giddings,
  ``Black holes, quantum information, and unitary evolution,''
Phys.\ Rev.\ D {\bf 85}, 124063 (2012).
[arXiv:1201.1037 [hep-th]].
}
\lref\Mathurrev{
  S.~D.~Mathur,
  ``Fuzzballs and the information paradox: A Summary and conjectures,''
[arXiv:0810.4525 [hep-th]].
}
\lref\SGerice{
  S.~B.~Giddings,
  ``The gravitational S-matrix: Erice lectures,''
[arXiv:1105.2036 [hep-th]].
}
\lref\GiPo{
  S.~B.~Giddings and R.~A.~Porto,
  ``The Gravitational S-matrix,''
Phys.\ Rev.\ D {\bf 81}, 025002 (2010).
[arXiv:0908.0004 [hep-th]].
}
\lref\Susstrouble{
  L.~Susskind,
  ``Trouble for remnants,''
[hep-th/9501106].
}
\lref\WABHIP{
  S.~B.~Giddings,
  ``Why aren't black holes infinitely produced?,''
Phys.\ Rev.\ D {\bf 51}, 6860 (1995).
[hep-th/9412159].
}
\lref\HaPr{
  P.~Hayden and J.~Preskill,
  ``Black holes as mirrors: Quantum information in random subsystems,''
JHEP {\bf 0709}, 120 (2007).
[arXiv:0708.4025 [hep-th]].
}
\lref\Susstrans{
  L.~Susskind,
  ``The Transfer of Entanglement: The Case for Firewalls,''
[arXiv:1210.2098 [hep-th]].
}
\lref\NLvC{
  S.~B.~Giddings,
  ``Nonlocality versus complementarity: A Conservative approach to the information problem,''
Class.\ Quant.\ Grav.\  {\bf 28}, 025002 (2011).
[arXiv:0911.3395 [hep-th]].
}
\lref\NVNLtwo{
  S.~B.~Giddings,
  ``Nonviolent information transfer from black holes: a field theory parameterization,''
[arXiv:1302.2613 [hep-th]].
}
\lref\GiSh{
  S.~B.~Giddings and Y.~Shi,
 ``Quantum information transfer and models for black hole mechanics,''
[arXiv:1205.4732 [hep-th]].
}
\lref\AMPSS{
  A.~Almheiri, D.~Marolf, J.~Polchinski, D.~Stanford and J.~Sully,
  ``An Apologia for Firewalls,''
[arXiv:1304.6483 [hep-th]].
}
\lref\Avery{
  S.~G.~Avery,
  ``Qubit Models of Black Hole Evaporation,''
JHEP {\bf 1301}, 176 (2013).
[arXiv:1109.2911 [hep-th]].
}
\lref\SGappear{S.~B.~Giddings, work in progress.}

\lref\UnWamine{
  W.~G.~Unruh and R.~M.~Wald,
  ``Acceleration Radiation and Generalized Second Law of Thermodynamics,''
Phys.\ Rev.\ D {\bf 25}, 942 (1982)\semi
``How to mine energy from a black hole," Gen. Rel. and Grav. {\bf 15}, 195 (1983).
}
\lref\Frolov{
  V.~P.~Frolov and D.~Fursaev,
  ``Mining energy from a black hole by strings,''
Phys.\ Rev.\ D {\bf 63}, 124010 (2001).
[hep-th/0012260]\semi
  V.~P.~Frolov,
 ``Cosmic strings and energy mining from black holes,''
Int.\ J.\ Mod.\ Phys.\ A {\bf 17}, 2673 (2002).
}
\lref\LawrenceSG{
  A.~E.~Lawrence and E.~J.~Martinec,
  ``Black hole evaporation along macroscopic strings,''
Phys.\ Rev.\ D {\bf 50}, 2680 (1994).
[hep-th/9312127].
}
\lref\BrownUN{
  A.~R.~Brown,
 ``Tensile Strength and the Mining of Black Holes,''
[arXiv:1207.3342 [gr-qc]].
}
\lref\LaLi{L.~D.~Landau and E.~M.~Lifshitz, {\sl Statistical Physics} (Part 1), Pergamon, 1980.}
\lref\GiMadS{
  S.~B.~Giddings and D.~Marolf,
  ``A Global picture of quantum de Sitter space,''
Phys.\ Rev.\ D {\bf 76}, 064023 (2007).
[arXiv:0705.1178 [hep-th]].
}
\lref\SeSu{
  Y.~Sekino and L.~Susskind,
  ``Fast Scramblers,''
JHEP {\bf 0810}, 065 (2008).
[arXiv:0808.2096 [hep-th]].
}
\lref\SuWi{
  L.~Susskind and E.~Witten,
  ``The Holographic bound in anti-de Sitter space,''
[hep-th/9805114].
}
\lref\FSS{
  S.~B.~Giddings,
  ``Flat space scattering and bulk locality in the AdS / CFT correspondence,''
Phys.\ Rev.\ D {\bf 61}, 106008 (2000).
[hep-th/9907129].
}
\lref\GGP{
  M.~Gary, S.~B.~Giddings and J.~Penedones,
  ``Local bulk S-matrix elements and CFT singularities,''
Phys.\ Rev.\ D {\bf 80}, 085005 (2009).
[arXiv:0903.4437 [hep-th]].
}
\lref\GaryMI{
  M.~Gary and S.~B.~Giddings,
  ``The Flat space S-matrix from the AdS/CFT correspondence?,''
Phys.\ Rev.\ D {\bf 80}, 046008 (2009).
[arXiv:0904.3544 [hep-th]].
}
\lref\GaGiAdS{
  M.~Gary and S.~B.~Giddings,
  ``Constraints on a fine-grained AdS/CFT correspondence,''
[arXiv:1106.3553 [hep-th]].
}
\lref\Isit{
  S.~B.~Giddings,
  ``Is string theory a theory of quantum gravity?,''
[arXiv:1105.6359 [hep-th]].
}
\lref\RaMu{
  J.~Rau and B.~Muller,
  ``From reversible quantum microdynamics to irreversible quantum transport,''
Phys.\ Rept.\  {\bf 272}, 1 (1996).
[nucl-th/9505009].
}
\lref\Sorkin{
  R.~D.~Sorkin,
  ``The statistical mechanics of black hole thermodynamics,''
[gr-qc/9705006].
}
\lref\GGS{
  D.~Garfinkle, S.~B.~Giddings and A.~Strominger,
  ``Entropy in black hole pair production,''
Phys.\ Rev.\ D {\bf 49}, 958 (1994).
[gr-qc/9306023].
}
\lref\DGGH{
  F.~Dowker, J.~P.~Gauntlett, S.~B.~Giddings and G.~T.~Horowitz,
  ``On pair creation of extremal black holes and Kaluza-Klein monopoles,''
Phys.\ Rev.\ D {\bf 50}, 2662 (1994).
[hep-th/9312172].
}
\lref\PageHP{
  D.~N.~Page,
  ``No-Bang Quantum State of the Cosmos,''
Class.\ Quant.\ Grav.\  {\bf 25}, 154011 (2008).
[arXiv:0707.2081 [hep-th]].
}
\lref\PageMQA{
  D.~N.~Page,
  ``Excluding Black Hole Firewalls with Extreme Cosmic Censorship,''
[arXiv:1306.0562 [hep-th]].
}
\lref\Brau{
  S.~L.~Braunstein, S.~Pirandola and K. \.Zyczkowski,
  ``Entangled black holes as ciphers of hidden information,''
Physical Review Letters 110, {\bf 101301} (2013).
[arXiv:0907.1190 [quant-ph]].
}
\lref\BaFi{
  T.~Banks, W.~Fischler,
  ``Space-like Singularities and Thermalization,''
[hep-th/0606260].
}
\lref\Sredent{
  M.~Srednicki,
  ``Entropy and area,''
Phys.\ Rev.\ Lett.\  {\bf 71}, 666 (1993).
[hep-th/9303048].
}
\lref\SuTh{
  L.~Susskind and L.~Thorlacius,
  ``Gedanken experiments involving black holes,''
Phys.\ Rev.\ D {\bf 49}, 966 (1994).
[hep-th/9308100].
}
\lref\BoLa{E. Bouchbinder and J.~S.~Langer, ``Nonequilibrium thermodynamics of driven amorphous materials. II.
Effective-temperature theory,"  Phys.\ Rev.\ E {\bf 80} 031132 (2009).}
\lref\WaldLiv{
  R.~M.~Wald,
  ``The thermodynamics of black holes,''
Living Rev.\ Rel.\  {\bf 4}, 6 (2001).
[gr-qc/9912119].
}
\lref\GaSt{
  D.~Garfinkle and A.~Strominger,
  ``Semiclassical Wheeler wormhole production,''
Phys.\ Lett.\ B {\bf 256}, 146 (1991)..
}
\lref\GiLi{
  S.~B.~Giddings and M.~Lippert,
  ``The Information paradox and the locality bound,''
Phys.\ Rev.\ D {\bf 69}, 124019 (2004).
[hep-th/0402073].
}
\lref\GiddingsSJ{
  S.~B.~Giddings,
  ``Black hole information, unitarity, and nonlocality,''
Phys.\ Rev.\ D {\bf 74}, 106005 (2006).
[hep-th/0605196].
}
\lref\SenBA{
  A.~Sen,
  ``Logarithmic Corrections to N=2 Black Hole Entropy: An Infrared Window into the Microstates,''
[arXiv:1108.3842 [hep-th]].
}
\lref\SenDW{
  A.~Sen,
  ``Logarithmic Corrections to Schwarzschild and Other Non-extremal Black Hole Entropy in Different Dimensions,''
[arXiv:1205.0971 [hep-th]].
}
\lref\HanadaEZ{
  M.~Hanada, Y.~Hyakutake, J.~Nishimura and S.~Takeuchi,
 ``Higher derivative corrections to black hole thermodynamics from supersymmetric matrix quantum mechanics,''
Phys.\ Rev.\ Lett.\  {\bf 102}, 191602 (2009).
[arXiv:0811.3102 [hep-th]].
}
\lref\JMR{
  T.~Jacobson, D.~Marolf and C.~Rovelli,
  ``Black hole entropy: Inside or out?,''
Int.\ J.\ Theor.\ Phys.\  {\bf 44}, 1807 (2005).
[hep-th/0501103].
}
\lref\Page{
  D.~N.~Page,
  ``Average entropy of a subsystem,''
Phys.\ Rev.\ Lett.\  {\bf 71}, 1291 (1993).
[gr-qc/9305007];
 ``Information in black hole radiation,''
  Phys.\ Rev.\ Lett.\  {\bf 71}, 3743 (1993)
  [arXiv:hep-th/9306083].
}
\lref\Mathurbit{
  S.~D.~Mathur,
  ``The Information paradox: A pedagogical introduction,''
Class.\ Quant.\ Grav.\  {\bf 26}, 224001 (2009).
[arXiv:0909.1038 [hep-th]].
}
\lref\BHIUNL{
  S.~B.~Giddings,
 ``Black hole information, unitarity, and nonlocality,''
Phys.\ Rev.\ D {\bf 74}, 106005 (2006).
[hep-th/0605196];
}
\lref\QBHB{
  S.~B.~Giddings,
  ``Quantization in black hole backgrounds,''
  Phys.\ Rev.\  D {\bf 76}, 064027 (2007)
  [arXiv:hep-th/0703116].
}
\lref\LPSTU{
  D.~A.~Lowe, J.~Polchinski, L.~Susskind, L.~Thorlacius and J.~Uglum,
 ``Black hole complementarity versus locality,''
Phys.\ Rev.\ D {\bf 52}, 6997 (1995).
[hep-th/9506138].
}
\Title{
\vbox{\baselineskip12pt
}}
{\vbox{\centerline{Statistical physics of black holes}\centerline{as quantum-mechanical systems}
}}
\centerline{{\ticp 
Steven B. Giddings\footnote{$^\ast$}{Email address: giddings@physics.ucsb.edu}  
} }
\centerline{\sl Department of Physics}
\centerline{\sl University of California}
\centerline{\sl Santa Barbara, CA 93106}
\vskip.40in
\centerline{\bf Abstract}

Some basic features of black-hole statistical mechanics are investigated, assuming that black holes respect the principles of quantum mechanics.  Care is needed in defining an entropy $\sbh$ corresponding to the number of microstates of a black hole, given that the black hole interacts with its surroundings.  
An open question is then the relationship between this entropy  and the Bekenstein-Hawking entropy $\sBH$.  
For a wide class of models with interactions needed to ensure unitary quantum evolution, these interactions produce extra energy flux beyond that predicted by Hawking.  Arguments are then presented that this results in  an entropy $\sbh$ that  is smaller than $\sBH$.  Correspondingly, in such scenarios equilibrium properties of black holes are modified.  We examine questions of consistency of such an inequality; if it is not consistent, that provides significant constraints on models for quantum-mechanical black hole evolution.

\vskip.3in
\Date{}

\newsec{Introduction}

A longstanding problem in quantum gravity is to characterize the quantum states of a black hole and their dynamics.  The statistical properties of these states and their corresponding thermodynamics should provide important guidance and constraints.  While there is an elegant treatment of black hole thermodynamics\foot{For a review and further references, see \WaldLiv.} based on the semiclassical description and associated with the Bekenstein-Hawking entropy $\sBH$, the same description leads to a violent clash with 
quantum-mechanical principles.  Thus, ultimately the semiclassical description must be an incomplete approximation.  Given this description's relation to thermodynamics, we can also ask whether or not a quantum black hole is well-described as a  thermal system with the Bekenstein-Hawking density of states.

In particular, if black hole disintegration respects unitarity, a challenge is to provide a description of the information transfer from the black hole interior that is necessary to restore quantum purity to the external state.  If semiclassical spacetime is still a good but not exact approximation to a large black hole, one expects radiation approximately as predicted by Hawking\refs{\Hawkrad}, but with some modifications that are necessary to accommodate the information transfer.  This transfer may specifically arise from additional processes yielding an increased flux of energy from a black hole, and this feature has been found in certain generic simple models for such processes\refs{\modelsone\modelstwo\GiSh\NVNL-\NVNLtwo}.  In turn, this raises the possibility that black hole decay into vacuum is not well-described as near-equilibrium decay at the Hawking temperature.  

For one  possible rough analogy, consider the Sun.  The temperature of the Sun's surface (more precisely, the photosphere) is approximately 6000 K, or 0.5 eV:  a nearby thermometer capable of withstanding such temperatures would measure this value.  But, at the same time, a much more sensitive detector could measure neutrinos streaming by with energies ranging from hundreds of keV to tens of MeV.  Clearly, while the solar atmosphere is in some respects approximately thermal, it is not a thermal equilibrium state at 6000 K.

Could a quantum black hole similarly have an approximate thermal description, for some purposes, yet not be accurately described as a thermal system with the Bekenstein-Hawking density of states?   If so, how is the black hole entropy defined and calculated, and what role does it play in thermodynamics?  Indeed, one might consider that the Bekenstein-Hawking entropy is not inevitably related to the number of black hole internal states, since it was regarded as an important black hole characteristic even in proposed scenarios where black holes destroy quantum information.  It may even be true that $\sBH$ and the Hawking temperature $T_H$ only characterize certain ``surface" properties and processes of a black hole -- like in the solar example.  A related question is to what extent black holes are star-like objects.  An extreme version of this is that of a massive remnant\BHMR, where the black hole horizon is replaced by an interface which is violent to infalling observers, or its variants such as fuzzballs\refs{\Mathurrev} or firewalls\refs{\AMPS}.  Alternatively, a black hole may appear to behave approximately semi-classically for most observers, who do not make sufficiently careful measurements\refs{\modelsone\modelstwo\GiSh\NVNL-\NVNLtwo}, with information leaking out due to relatively small effects.\foot{For a loose analogy, consider helium diffusively leaking from a balloon.}

Regardless of the answers to these questions, in general it is important to understand  the respective roles of the different possible entropies characterizing a black hole.  We begin by discussing the question of what one might mean by the careful definition of an entropy $\sbh$ corresponding to the number of black hole states, given that black hole interactions with an environment are generically present.  Then, we turn to questions of constraining and characterizing this entropy.  One such question regards how the $\sim \exp\{\sbh\}$ states can be produced.  This question is followed by discussion of the implications of unitary evolution for $\sbh$, and the relation between $\sbh$ and the entanglement entropy $\svn$ of a black hole with its environment.  Following the preceding discussion, we explore the possibility that $\sbh<\sBH$, as indicated by related arguments.  This, and the question of how a black hole returns information to its environment, also has possible implications for the nature of the equilibrium state of a black hole with a thermal system.  An important question which we then briefly address is whether there is any fundamental inconsistency in the statement $\sbh<\sBH$, either internally, or with known physical facts.  

For the purposes of this discussion, we assume that a black hole interacting with an environment is well-described within the framework of quantum mechanics.  We are not yet able to address the important question of the nature of the black hole's microstates, or their detailed evolution.  But, basic key features of the statistical mechanics of black holes are expected to not necessarily depend on such details, given the large number of black hole states.  An important question is that of finding the constraints that consistency places on such a description, and in particular on the form of the information transfer needed for unitarity, and on the related presence of extra flux from black holes.

\newsec{Basics of a statistical description}

Begin by considering some essential aspects of a statistical description for $D$-dimensional black holes.  If a black hole 
interacting with its environment is a  system governed by the principles of quantum mechanics -- a basic assumption of this paper -- we then expect that an important quantity in a statistical description of the quantum black hole is the number of its states, $\Omega^{bh}_{\delta M}(M)$, in a range of black hole masses $(M, M+\delta M)$. (More generally we may wish to also consider dependence on macroscopic parameters such as angular momentum $J$ or charge(s) $Q$.)  We moreover assume that the environment of the black hole, and the interactions with it, are approximately described by local quantum field theory (LQFT), plus corrections necessary to restore unitarity.   The scenario described in \refs{\modelsone\modelstwo\GiSh\NVNL-\NVNLtwo} assumes that these corrections are small, in an appropriate sense.

Some care is needed, though, in defining $\Omega^{bh}_{\delta M}(M)$, since a black hole will interact with its environment.  In general, one cannot turn off both absorption of matter and its emission ({\it e.g.} in Hawking radiation), even for a black hole placed in vacuum.  Thus a black hole is not intrinsically isolated -- quantum states are jointly those of the black hole and environment together.

However, we expect there to be approximations where black hole states can be meaningfully counted.  For example, consider a black hole in vacuo. The time required for a substantially non-extremal ($M\gg J, Q$) black hole state to decay, emitting a Hawking quantum,\foot{More rapid decay will be considered later in this paper; in its presence, the energy range in the following discussion can be adjusted appropriately} is $\sim R(M)$, where $R(M)$ is the Schwarzschild radius.\foot{In $D$ dimensions, $R(M)\propto M^{1/(D-3)}$.  Most conclusions of this paper are expected to apply for $D\geq 4$, though the specific example $D=4$ will for some purposes be considered.}  Therefore, at shorter times we expect to have a precise notion of black hole quantum states.  Equivalently, this decay time scale means 
black hole quantum states have widths $\Gamma(M)\sim 1/R(M)$.  Correspondingly, we should only define the black hole density of states for intervals with $\delta M\roughly> 1/R(M)$.  Then we expect the number of states to be proportional to $\delta M$,
\eqn\dos{\Omega^{bh}_{\delta M}(M) = \omega^{bh}(M) \delta M\ .}
Another basic assumption of this paper is thus that there is such a well-defined density of black hole states; this then should capture important features of  black holes both in  vacuum, and in equilibrium with a thermal environment.

The quantity $\omega^{bh}(M)$ is also an important one for information-theoretic properties of black holes, which are of course closely related to thermodynamic properties.  In particular, the number of quantum states provides an upper bound on the amount of entanglement a black hole can have with its environment.  This, in turn, constrains information transfer to and from the black hole; if a black hole shrinks and the amount of information it can contain decreases, and evolution is quantum-mechanical, information must transfer from the black hole to its environment.

\newsec{Black hole density of states: constraints and expectations}

One way to infer properties of the basic theory describing black hole quantum states is to examine constraints on and expectations for the density of states. A black hole entropy may be defined via\foot{More generally, for a non-Schwarzschild black hole, one may wish to \eg\ replace $R(M)\rightarrow 1/\Gamma(M)$.}
\eqn\entdef{\Omega^{bh}_{\delta M}(M) = \omega^{bh}(M) \delta M = e^{\sbh(M)} R(M) \delta M\ .}
A longstanding and widespread expectation is that $\sbh(M)=\sBH(M)$, where $\sBH$ is the Bekenstein-Hawking entropy, but we should investigate what other constraints tell us about $\sbh$.  

Any $\sbh$ remotely approaching $\sBH$ would mean that the black hole internal states are extremely closely spaced, with spacing $\Delta M\sim \exp\{-\sbh\}/R(M)$.  While such quantum states are narrow, at least given the semiclassical estimate $\Gamma(M)/M \sim 1/\sBH\ll 1$, their width is far greater than their spacing.  Put differently, isolating a single quantum state is expected to take a time $\Delta t\sim R(M)\exp\{\sbh\}$, far larger than the decay time of the black hole.  This further motivates the statistical approach to describing black hole states.

An important question is how to excite all the states described by $\Omega^{bh}_{\delta M}(M)$.  For example, matter collapsing to form a black hole has been estimated\refs{\thooholo} to carry entropy $\sim \sBH^{3/4}\ll \sBH$.  One may alternately collide particles in a pure quantum state to make a black hole.\foot{For a review and further references, see \refs{\SGerice}.}  In this latter case, radiation, gravitational and otherwise, is typically emitted.  But, this at most provides an entropy $\sim \epsilon \sBH$, with $\epsilon\ll1$, and so does not provide a means of exciting $\Omega^{bh}_{\delta M}$ distinct states.\foot{Note that a collision from a pure state will in general produce radiation states entangled with the black hole states, but this entanglement is limited by the number of quantum radiation states.  For further discussion, see \refs{\GiPo}.}

However, if the semiclassical description is to be trusted, the Hawking process does provide a means to excite the $\Omega^{bh}_{\delta M}(M)$ states of a black hole with mass $M$: one begins with a black hole with mass $2M$, and allows it to Hawking radiate to mass $M$, in the process producing entanglement entropy $\sim \sBH$ between the black hole and radiation states\GiPo.  Thus, if we project on one of the $\sim \exp\{\sBH\}$ states of the radiation, we also project on a definite corresponding state of the black hole.\foot{In the terminology of information theory, the radiation states provide a purification of the black hole state.   This discussion ignores subtleties of ``projection" vs. quantum measurement.}  An alternate process 
 is to begin with a black hole of mass $M$, and 
inject information in the form of ingoing quanta with a flux of energy matching the outgoing Hawking flux.
On a time scale comparable to the decay time just described, $t\sim R(M)\sBH$, the semiclassical approximation tells us this can populate $\sim \exp\{\sBH\}$ distinct states of the black hole. Yet another such semiclassical process is black hole pair production\refs{\GaSt\GGS-\DGGH}.

Of course, quantum mechanics tells us that the semiclassical description must ultimately fail.  The preceding discussion could have been extended to show that a black hole of initial mass $M_0$ has $\sim \exp\{\sBH(M_0)\}$ states even after evaporating to the Planck mass.  The result, for increasing $M_0$, is an unboundedly large number of species of planckian black hole remnants, which has been argued to yield instability to catastrophic remnant production that is inconsistent with observation, or implies other inconsistencies.\foot{See {\it e.g.} \refs{\WABHIP,\Susstrouble}.}  

By these arguments, the true $\sbh$ must be less than that calculated in such a semiclassical approximation.  
To preserve quantum mechanics and avoid remnant instabilities, new effects beyond the semiclassical description of Hawking radiation must 1) give us finite $\sbh$ and 2) describe quantum information transfer\foot{Such information transfer may be characterized in terms of transfer of the entanglement of the black hole degrees of freedom with other degrees of freedom; see \refs{\HaPr,\GiSh,\Susstrans}.} out of a shrinking black hole before it reaches the Planck size.  A critical question is to describe the new physics responsible for these effects.   Any constraints on this physics furnish important clues.  

While in order to preserve quantum mechanics, such new physics must apparently be operative in black holes and their immediate surroundings,  we also find that local quantum field theory(LQFT) in semiclassical spacetime gives an excellent description of all effects observed in nature so far. For this reason, it seems plausible that LQFT gives a good approximation for the environment of a black hole outside its immediate atmosphere, the latter extending outward to a few times $R$.  So, the simplest and most conservative possibility appears to be that in the correct description of quantum black holes, information transfers from a black hole into its atmosphere, violating the restrictions of LQFT on superluminal information transfer\refs{\BHMR,\GiLi\GiddingsSJ-\NLvC,\modelsone\modelstwo-\GiSh,\AMPS,\NVNL,\NVNLtwo}.  Such an extreme measure seems a necessary response to the difficulty of restoring unitary evolution to save quantum mechanics in the presence of black holes.

Another important entropy for characterizing black hole interactions is that of von Neumann.  If we suppose that a black hole and its surroundings are in a pure quantum state $|\psi\rangle$, {\it e.g.} formed in a pure-state collision,  tracing over the black hole or internal degrees of freedom,
\eqn\densmats{ \rho_{ext}= {\rm Tr}_{bh}|\psi\rangle\langle\psi|\quad ,\quad \rho_{bh}= {\rm Tr}_{ext}|\psi\rangle\langle\psi| }
gives density matrices for the exterior or black hole states.  Then, the von Neumann entropy of either is given by\foot{There are important subtleties in these expressions depending on the cutoff used to separate the black hole and exterior states, but any subdivision that is sharp on a scale $\roughly<R$ is expected to ultimately capture most of the entropy of the radiation and the black hole.  
}
\eqn\Svn{\svn= -{\rm Tr} \rho_{ext}\ln \rho_{ext} = -{\rm Tr} \rho_{bh}\ln \rho_{bh}\ .}
Following the above discussion, through Hawking radiation $\svn$ reaches a size $\sim \sBH$ at time $t\sim R \sBH$.  On the other hand,  absence of infinite planckian remnant degeneracy  implies $\svn$ decreases to a finite fixed value, which we assume to be zero, after the final decay of a black hole.  The value of $\sbh$, defined above, plays a key role in the evolution of $\svn$, since unitary evolution implies $\svn\leq \sbh$.  An upper bound is also given by the entropy that the radiation can carry,\foot{A useful measure of the available radiation entropy is simply the usual coarse-grained entropy of the radiation.} resulting in 
\eqn\svnbd{\svn\leq {\rm Min} (\sbh, S_{rad})\ .}

\Ifig{\Fig\entropyc}{The curves of radiation entropy $S_{rad}$ and Bekenstein-Hawking entropy $\sBH$, plotted against the ratio of radiation energy to initial mass, $E/M_0$,  bound the von-Neumann entropy \Svn\ from above.  The Page curve, to an excellent approximation, follows the minimum of these two curves, and defines the Page time $\tpage$.  If, as argued subsequently, $\sbh<\sBH$, the entropy $\Svn$ begins to fall sooner, resulting in a curve like the lower one.}{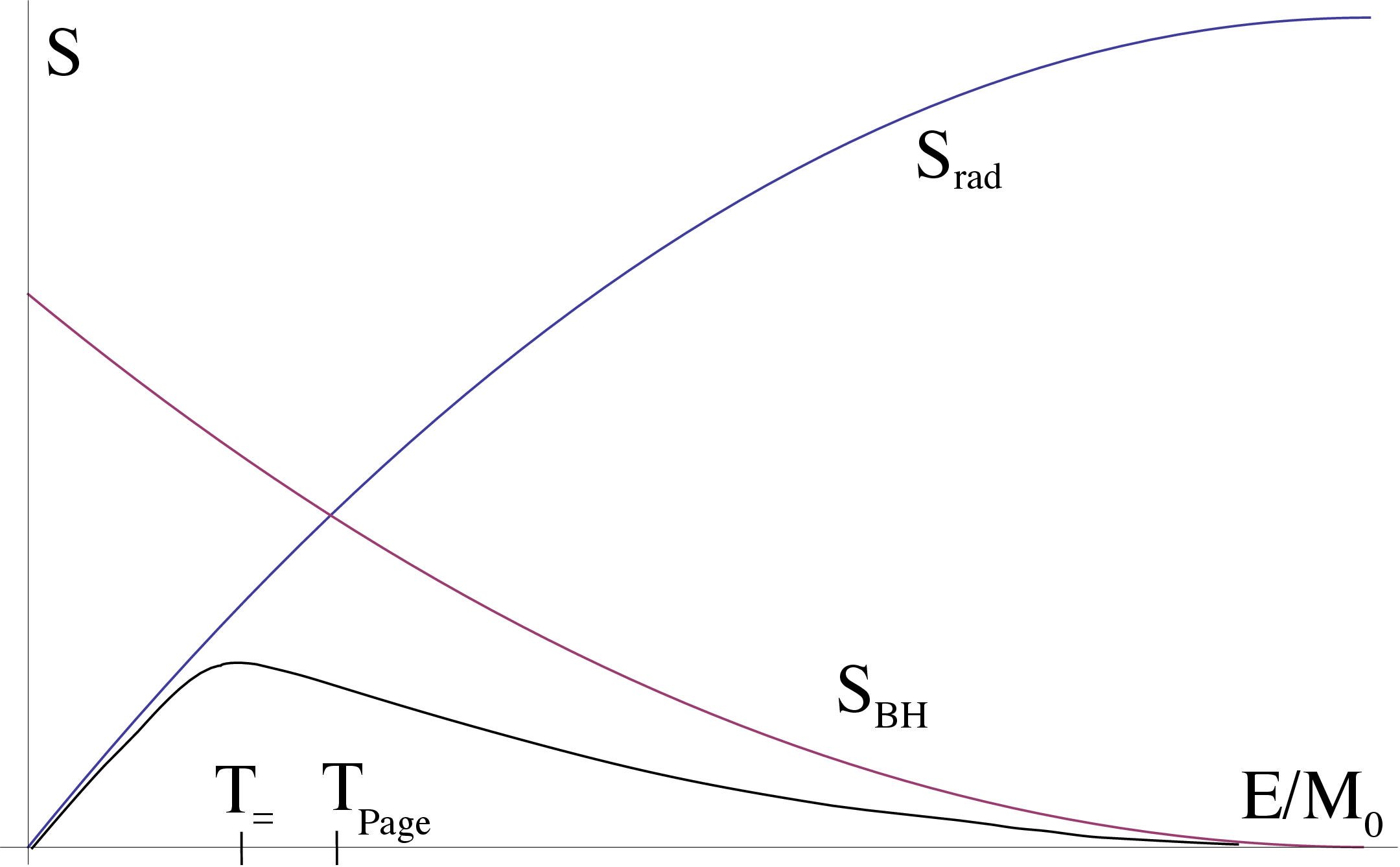}{5}

Without providing any fine-grained description, Page\refs{\Page} suggested an important constraint on  unitary evolution, assuming that $\sbh=\sBH$. Specifically, by also assuming that the evolution coupling the black hole and exterior radiation systems is well approximated as random unitary evolution, he showed that $\svn$ {\it saturates} the bound \svnbd\ with $\sbh=\sBH$, to an excellent approximation.  
If this is the correct picture, $\svn$ begins to decline where $\sBH=S_{rad}$; the corresponding time is the Page time, $\tpage$.  (See \entropyc.)

The problem is to find a fine-grained dynamics respecting these or related constraints.  This problem has been particularly well-illustrated by the confusions surrounding the possibility that such effects alter correlations at the horizon sufficiently to make it singular\refs{\BHIUNL\Mathurbit-\Brau,\modelsone,\modelstwo}, exemplified in the ``firewall" debate\refs{\AMPS,\AMPSS}.  While the firewall picture apparently assumes the kind of superluminal information transfer that we have described, in order for the information to transfer out of a black hole, it assumes a particularly violent transfer  to modes just at the horizon, resulting in a destructive interface that replaces the horizon -- just as in the earlier massive remnant scenario\BHMR.  

An important question is whether unitarity can be preserved through a more non-violent form of information transfer, as proposed in \refs{\NLvC,\modelsone\modelstwo\GiSh\NVNL-\NVNLtwo}.  In particular, we would like to investigate whether the assumption that physics near the horizon is for many purposes well-described by the approximate semiclassical geometry is consistent with quantum mechanics.  If not, our semiclassical picture of black holes changes drastically. 

Quantum fields fluctuating about the semiclassical geometry produce Hawking radiation, so if this picture is approximately correct, we expect black holes to emit similarly in a complete description.  The question is how they can do so, while at the same time emitting the required information.  Simple models for this are described in \refs{\modelsone,\Avery,\modelstwo,\GiSh,\AMPS,\NVNL,\NVNLtwo,\AMPSS}.  In particular, as discussed in \refs{\modelstwo\NVNL-\NVNLtwo}, a generic property of the models considered is the prediction of extra energy flux, beyond the value predicted by Hawking.

A simple way to think of this is that if the Hawking effect is present, the black hole emits energy while increasing its entanglement.  To decrease the entanglement, extra excitations have to be emitted.  The models of \refs{\modelsone,\modelstwo,\NVNL,\NVNLtwo} typically have this property.  There are two known kinds of exception.  One is described in \AMPS: external Hawking modes are exchanged with information-bearing modes inside the black hole, sequestering the entanglement that would have been produced, and simultaneously emitting the necessary information.  Models such as this requiring two-way non-LQFT transfer will not be considered further.  The second is model two of  \refs{\modelsone,\modelstwo}.  In this model, two qubits of information are emitted in the Hawking modes, through encoding in an alteration of the Hawking state.  In the infinite-temperature limit or zero-frequency limit, where $\exp\{-\beta \omega\}=1$, it was found that such evolution produces an energy flux equal to Hawking's.  However, consideration of the finite temperature case shows that the flux exceeds this rate\refs{\SGappear}.  Moreover, investigation of information transfer that can be modeled as a modification of effective field theory \NVNLtwo\ reveals criteria for extra flux to be avoided, and that such criteria are not easily satisfied\refs{\SGappear}.  

Thus, there are two possibilities.  The first is that special ``Hawking-like" models exist which both transfer the needed information out and match the Hawking energy flux, unexpectedly evading the stringent constraints on doing so.  The second is that the correct dynamics indeed predicts additional energy flux out of a black hole.  The latter will be an important scenario to investigate in the remainder of the paper.  This has significant consequences for our discussion of statistical and thermal properties of black holes.

Specifically, if $E$ denotes the energy radiated from the black hole, the condition
\eqn\bigerrad{{dE\over dt}> {dE\over dt}_{\Big\vert_{\rm Hawking} }}
implies, under fairly general assumptions, $\sbh<\sBH$.  

The argument for this is relatively simple, and was essentially given in \NVNL\ (see also \AMPSS).  Specifically, if like Page we assume that the internal degrees of freedom are ultimately mixed with the emitted degrees of freedom by what looks like a random unitary, then $\svn$ will saturate the bound \svnbd.  If the energy flux exceeds Hawking's value, then as $\svn$ declines, 
\eqn\entropineq{{d\svn\over dE} > {d\sBH\over dE}\ }
(note both quantities are negative).  But, if at any time $\svn=\sBH$, this is inconsistent with the bound \svnbd\ -- black holes become ``overfull," and once they fully disintegrate, we are returned to the inconsistencies of planckian remnants. This can only be avoided while respecting \entropineq\ if $\sbh<\sBH$
and results in a curve like that illustrated in \entropyc.  Notice that this means $\svn$ begins to decrease -- and thus information is emitted -- {\it before} the Page time $\tpage$.  The random-unitary assumption could be relaxed -- for example one could consider evolution that reemits the entanglement carried by the inside Hawking excitations at a time $10 R$ after their outside partners left the black hole, resulting in a much lower $\svn$, or one could imagine some evolution intermediate between these extremes.  But these would essentially imply that the degrees of freedom counted by $\sbh$ are irrelevant since they do not all become excited, which is effectively the same result.  Earlier information transfer results in smaller $\Omega^{bh}_{\delta M}(M)$.

Specifically, in this scenario, black hole disintegration is {\it not} correctly characterized as equilibrium emission from a black hole with entropy $\sBH$.  

In some ways, this result seems similar to the stellar example discussed in the introduction.  The black hole emits Hawking radiation that is approximately thermal, resulting from usual LQFT processes near the horizon  -- somewhat analogous to stellar emission from the photosphere.  However, there is a process of emission of extra excitations that is not necessary thermal, and originates in the quantum dynamics of the black hole interior -- somewhat analogous to stellar emission of neutrinos.\foot{Indeed, to develop the solar analogy further, imagine that the weak interactions had been first discovered in the context of providing a suitable theory of the Sun.  As noted, the Sun's atmosphere is approximately  in local thermal equilibrium, for some purposes.  But, an essential feature of solar physics is that there are other processes in operation.  These are necessary both to conserve energy, and to conserve lepton number. And, if one does not account for the escaping neutrinos, solar physics would appear to violate both of these important conservation principles.  In the present context, we consider the possibility that there are similarly additional processes present besides the approximately thermal process of Hawking radiation.  These processes are, like weak interactions, necessary to satisfy an important  conservation principle -- here that of quantum information.  And, they result in an additional flux of energy -- that may be hard to see by conventional means -- beyond that of the thermal process.
}

We have thus found that the constraints of unitary evolution, and that black holes are well-approximated by their semiclassical description, indicate $\sbh(M)<\sBH(M)$.  Therefore, an important question is whether there is any basic inconsistency in this inequality.  If so, that would imply either a loophole in the preceding discussion (possibly via some special Hawking-like unitary evolution), or that the semiclassical picture of black holes fails, producing star-like massive remnants, with big departures from black-hole behavior, such as firewalls.  (Of course, the latter threaten not just to abandon the semiclassical picture of black holes, but {\it also} black hole thermodynamics.)  

One might note that the story of black hole  thermodynamics has been developed at great length without a microscopic calculation of $\sbh$; perhaps $\sBH$ is just an effective quantity, describing the ``surface" dynamics of Hawking radiation.\foot{For related discussion, see \refs{\Sorkin,\JMR}; the latter considers $\sbh>\sBH$.}  But, certainly $\sbh<\sBH$ may imply other surprises, and at the least indicates significant modification of the ultimate thermodynamics of black holes.

A comment also can be added regarding the process of black hole mining\refs{\UnWamine\Frolov\LawrenceSG-\BrownUN}, in which introduction of a cosmic string or other mining apparatus increases the rate of energy flow out of a black hole.  In order to avoid an overfull black hole, this means that the rate of information transfer out of the black hole must increase in a commensurate fashion.  The effective source models of \refs{\NVNLtwo} give one way to achieve this commensurate increase in information transfer with energy:  both arise from extra channels for outward flow of degrees of freedom being opened by the mining apparatus.  Correspondingly, $\Delta \svn/\Delta E$ could remain of approximately constant size independent of mining.  This, in turn, suggests that the curve $\svn(E)$ -- illustrated in \entropyc\ -- may not be significantly changed by the presence of mining.

\newsec{Black holes and equilibrium}

\subsec{The question of equilibration of subsystems}

An important question is to what extent and in what circumstances a black hole can be treated as a system in equilibrium. Consider first a black hole evaporating into vacuum.  The black hole is not a closed system, but might be thought of as quasi-closed\refs{\LaLi} since the interaction with the environment is relatively weak.\foot{Note that even at times larger than the Hawking emission time, $t\sim R$, for some purposes the black hole is effectively quasi-closed, as only a relatively small number of surface/atmosphere states are expected to interact with the exterior on this timescale\refs{\modelstwo}.}  A conventional definition is that in equilibrium, all microstates are equally likely.  Lacking a precise description of internal black hole states and dynamics, this is hard to test.  But, at a less refined level we can consider the question of equilibration of subsystems of a black hole, and address the question by investigating the expected form of interactions between these subsystems.  So, a key question is how to describe such subsystems.

\Ifig{\Fig\bhfactors}{Schematic of the different ``regions" of a black hole, corresponding to different factors of the total Hilbert space.}{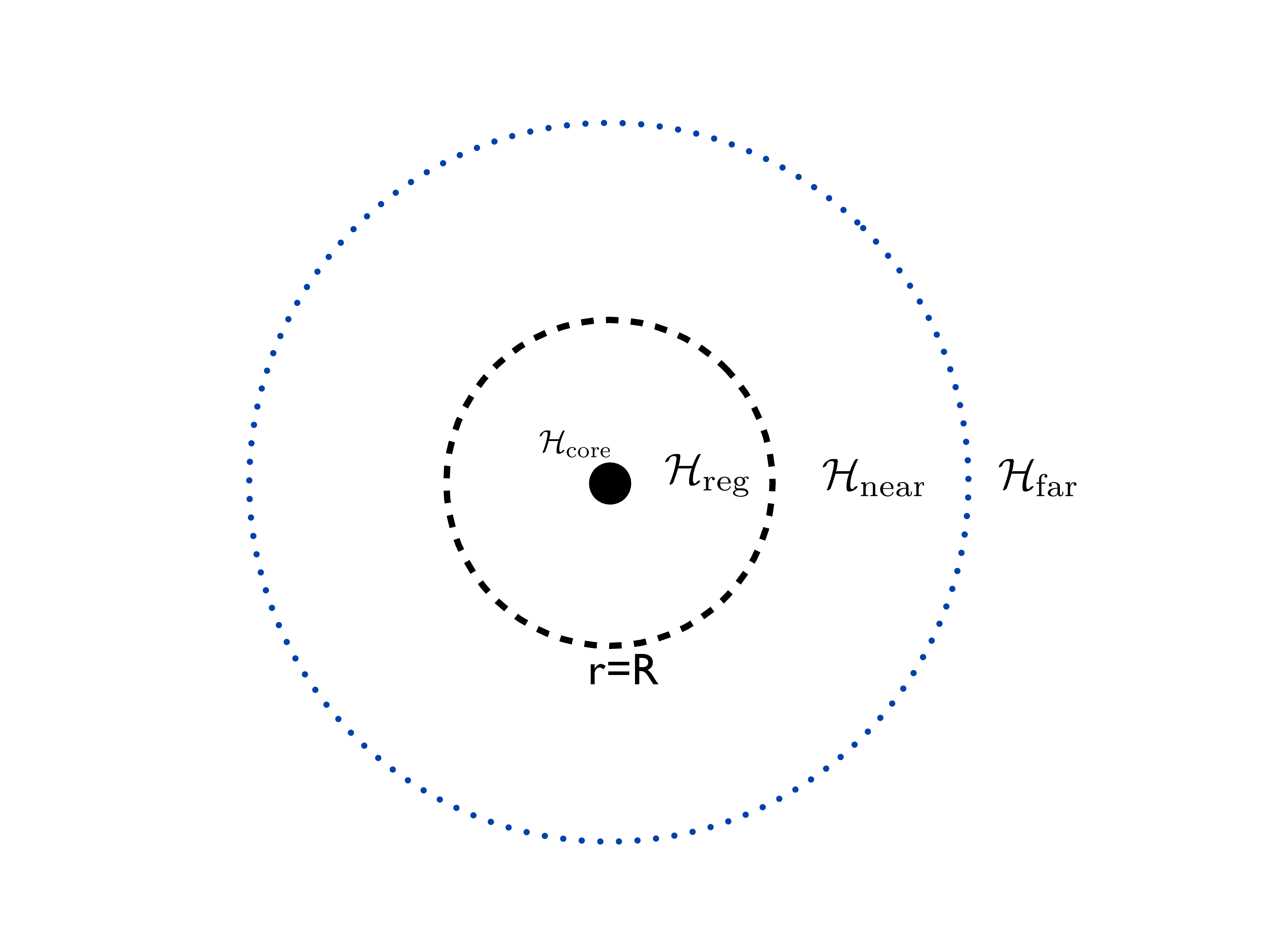}{6}

Following \refs{\modelstwo,\GiSh}  (see also \Susstrans), we assume that the states of the black hole plus environment are contained\foot{Note that not all states of the product may be realizable physical states, as discussed in \refs{\GiMadS\PageHP-\PageMQA}.} in a product Hilbert space 
\eqn\hilbprod{\calh=\hbh\otimes\hnear\otimes\hfar}
where $\hbh$ are the states of the black hole ``interior," $\hnear$ those of the immediate atmosphere, and $\hfar$ those further out.  One may envision an additional possible refinement\refs{\NVNLtwo} and for a large black hole write $\hbh\subset\hcore\otimes\hreg$, where $\hcore$ describes states in the core, ``strong-curvature" region of the black hole, semiclassically at $r\ll R$, and $\hreg$ describes states in the remaining weakly-curved region of the black hole interior, where one expects ordinary observers could make measurements.  

Without a detailed treatment of unitary black hole dynamics, one can describe important features of it by characterizing the information transfer among these subsystems, and the relevant timescales.  These characteristics are in turn important for addressing equilibration of black holes.

To begin with, consider the dynamics predicted by LQFT evolution on the semiclassical background of the evaporating black hole.  This evolution propagates information from $\hnear$ to $\hbh$, but, by locality of LQFT, not from $\hbh$ to $\hnear$.  Moreover, given a  slicing, one may give a refined description of the interior evolution.  One example is via the nice slices\LPSTU\ given in\refs{\modelstwo}
\eqn\NSeq{X^+(X^-+ e^{kT} X^+)=R_c^2\ ,}
where $X^\pm$ are Kruskal coordinates, $T$ is a parameter labeling the slices, and $k$ is a constant;
these asymptote to a constant $r=r_c(R_c)\ll R$ inside the black hole.  The states on the portion $R>r\gg r_c$ give $\hreg$; curvature here is weak and evolution is expected to be well-described by LQFT, which predicts that all excitations evolve to decreasing $r$.  States near $r=r_c$ are those of $\hcore$.  Since the lapse $N$ vanishes at $r=r_c$, the evolution predicted by LQFT freezes there\refs{\QBHB}.  Alternately, one could consider a ``natural" slicing\refs{\NLvC} that extends to $r<r_c$ and reaches arbitrarily strong curvature; here LQFT manifestly fails.\foot{Ref.~\NVNL\ proposes that these descriptions may in fact be gauge equivalent, and also possibly gauge equivalent to a description based on a Schwarzschild slicing.}  
Evolution also produces paired Hawking excitations in $\hreg\otimes\hnear$, increasing the entanglement between the black hole and its environment.  The inside excitations evolve inward to $\hcore$, and aside from a reflected part, the outside excitations propagate into $\hfar$.  

As we have described, this picture needs modification to restore unitary evolution.  In particular, it gives an unboundedly large $\Omega^{bh}(M)$ if one starts with an arbitrarily large initial black hole that then evaporates to a given size.  Instead, apparently unitary black hole dynamics requires  $\sbh\leq\sBH$.  Correspondingly, unitary evolution must include processes beyond the previous LQFT description, that transfer information from $\hbh$ to $\hnear$.\foot{One could also consider transfer to $\hfar$, representing nonlocality on scales large as compared to $R$.  However, this paper will restrict attention to the more conservative possibility that any requisite modification of LQFT operates primarily within a scale set by $R$.  Note in particular that there are a many modes available to carry information at very low frequencies, so which could be excited for a given finite temperature, but that exciting these modes requires longer-scale nonlocality.}

\subsec{Processes and timescales}

Here, we have assumed that most of the 
$\Omega^{bh}$ states are contained in $\hbh$ (if the semiclassical approximation is a good guide we moreover expect them to lie in $\hcore$).\foot{Of course one must carefully define the Hilbert-space decomposition with an appropriate regularization.  For example, a planckian cutoff yields\Sredent\ an entropy $\sim \sBH$ associated with a surface like the black hole horizon; however, the corresponding modes at such short scales are not typically excited.}
The unitary evolution that goes beyond the previous LQFT picture may be characterized by different processes acting on different timescales\refs{\modelstwo\GiSh-\NVNL}.  

\subsubsec{Scrambling}  

One such possible process is {\it  scrambling} of degrees of freedom in $\hbh$, with characteristic timescale $\tsc$.  LQFT nice-slice evolution predicts $\tsc=\infty$ (evolution freezes, never scrambling), but evolution on natural slices leads to the expectation\BaFi\ of a gauge-equivalent\NVNL\ description with $\tsc\sim R$.  If the semiclassical picture is a good guide, we expect this scrambling to primarily act on $\hcore$.  

\subsubsec{Transfer}  

A second process is {\it transfer} of information from $\hbh$ to $\hnear$.\foot{This may be sharply defined in terms of transfer of entanglement with an auxiliary system \refs{\HaPr,\GiSh,\Susstrans}.}  (In a refined description, this could be from $\hcore$ to $\hnear$ as well as to $\hreg$.)  When described with respect to the approximate semiclassical geometry, this transfer is superluminal, which would be forbidden in LQFT.  We call the characteristic timescale on which this operates $\tx$.

These timescales are relevant for equilibration:  $\tsc$ is the relaxation time scale for mixing of internal black hole degrees of freedom, and $\tx$ governs a black hole's equilibration with its environment.

\subsubsec{Discussion of timescales}

While the semiclassical analysis predicts\foot{This statement is made more sharp by considering a black hole fed energy at a rate matching that of outgoing Hawking radiation} $\tsc=\tx=\infty$, unitarity requires $\tx< \tpage$, and, if $\sbh<\sBH$, moreover requires $\tx$ to be less than or equal to  the time $T_=$ where $\sbh=S_{rad}$.  $\tx$ could be even smaller; if it is, and if dynamics is sufficiently close to random, $\svn$ does not begin to decline at such an earlier time, despite the transfer of entanglement.  Another key question is the dynamics predicting these timescales.

The fast scrambling conjecture\refs{\HaPr,\SeSu} states that $\tsc\sim \tx\sim R\ln R$.  This is motivated by the fact that the classical geometry of a black hole relaxes on the timescale $R\ln R$, which is suggestive of equilibration.  However, such a short timescale is a {\it maximal}\refs{\SuTh} departure from the semiclassical prediction of the timescale relevant for information transfer.  An alternate possibility that seems reasonable is that only local equilibration of the near-horizon atmosphere takes place on the time scale $R\ln R$, and that this scale does not characterize relevant information transfer rates from the black hole, and thus equilibration of black hole degrees of freedom with an external system.

If $\tx\gg R\ln R$, and even approaches $R\sbh$,  then early radiation at the Hawking temperature does not necessarily arise from equilibrium of the complete set of black hole degrees of freedom.\foot{We have noted that in the semiclassical approximation $\tsc$ is gauge dependent;  while one might expect\refs{\BaFi} the `physical' result that $\tsc\sim R$, we do not need this result.}  Moreover, when information transfer becomes important at $\tx$, this is not necessarily a thermal or equilibrium process.\foot{Compare neutrino emission from the Sun.}  Thus, black hole disintegration into vacuum may involve multiple processes, with different relaxation times, that are not necessarily simultaneously in equilibrium.

\subsec{Black hole equilibrium with a thermal environment}

At times longer than the relevant relaxation times, particularly $\tx$, we expect that a black hole in contact with a thermal environment could come into  equilibrium with that environment.  The preceding considerations have potentially important consequences for ultimate nature of the resulting equilibrium state.  By way of motivation, we note that a truly equilibrium configuration of a star has significantly different characteristics than that of the sun.

In view of the preceding discussion of subsystem decompositions, we can write the density of states for a black hole in contact with such an environment as
\eqn\totDOS{\omega(E_0) = \int dE\, \omega^{\overline{\rm bh}}(E)\, \omega^{\rm far} (E_0-E)\ ,}
where we denote the combined systems (bh+near) by $\overline{\rm bh}$.

A useful way to construct a gravitational box of thermal radiation is using anti-de Sitter space, as has been well-explored in the literature.  We first recall the semiclassical picture, based on equilibration with the Hawking radiation.  A small black hole, with radius $R\ll \rads$, cannot be in stable equilibrium, due to the negative specific heat of the black hole.  This can be regarded as a failure to maintain a detailed balance condition:  if the black hole absorbs energy $\Delta E$, it decreases its Hawking temperature and emits less, with unchanged incident energy.  In a large thermal bath, the black hole would keep growing.  

However, above the threshold size
\eqn\BHthresh{R\sim \rads^{3/5}}
a black hole can come into stable equilibrium with a  thermal bath; this relies on a finite volume effect where a black hole that radiates an energy $\Delta E$, increasing its Hawking temperature, also consequently increases the Hawking temperature of the thermal bath.  For an AdS box of size $R\roughly>\rads$, the black hole has size comparable to that of the box. Such ``large" black holes essentially dominate the thermal bath; any radiation emitted reflects off the walls and is reabsorbed on a timescale $\sim R$.    

This semiclassical equilibrium results from the Hawking radiation rate, but we are investigating  a scenario where unitary black hole disintegration emits energy at a higher rate, \bigerrad.  If so, one expects that a black hole would equilibrate at a higher temperature than the Hawking temperature:  in order to achieve detailed balance with the higher outgoing flux, the surrounding thermal environment must be raised to a higher temperature.\foot{This assumes that the interactions responsible for the extra outward flux do not also significantly increase absorption.  At frequencies $\omega\roughly>R$, a black hole is a near perfect absorber, while emitting negligibly.  The small interactions necessary to achieve the desired emission thus do not necessarily give significant fractional corrections to absorption.  For an analogy, the small diffusion effects of helium out of a balloon do not imply significant corrections to the scattering of outside helium atoms from the balloon's surface.}  A higher equilibrium temperature $T>T_H$ also corresponds to a lower entropy $\sbh$, through the standard thermodynamic relation
\eqn\tempvsent{{1\over T} = {\partial \sbh\over\partial M} \simeq {\partial \ln \Omega^{bh}\over \partial M} \ .}
For a benchmark estimate, consider, {\it e.g.} an entropy flux twice that of Hawking.  Basing the estimate on the Stefan-Boltzmann law, this means
$T=2^{1/3} T_H$, suggesting a decrease of $\sbh$ by $1/2^{2/3}$.

\newsec{Exploring further constraints on $\sbh$}

This paper has proposed the possibility that the entropy $\sbh$ characterizing the number of microstates of a black hole is smaller than the Bekenstein-Hawking entropy:  
$\sbh<\sBH$.  An obvious question is whether this presents any logical contradiction, or contradicts known results.  If one were to find such a contradiction, or otherwise show $\sbh=\sBH$, then going back through the logic described above provides significant constraints on scenarios for unitary black hole disintegration, and in particular suggests that such scenarios cannot result in extra energy flux beyond Hawking's.  This question of counting black hole states of a highly non-extremal black hole has arisen in several physical contexts, which should be assessed.

\subsec{AdS/CFT, and string/brane microstates}

An actual calculation of the entropy of a black hole in AdS, through AdS/CFT, first of all requires a calculation in the strongly coupled regime $g^2N\gg1$, of the field theory.  While suggestive estimates have been given in \refs{\SuWi}, actual strong-coupled calculations are not presently possible. A simpler situation to explore is that of $2+1$-dimensional AdS, with a boundary $1+1$-dimensional CFT, and entropy calculations through the Cardy formula.  However, gravity and quantum black holes in $2+1$ dimensions have significantly different properties than their higher-dimensional relatives, and so it is also not clear to what extent such a calculation could be extrapolated or extended to the higher-dimensional case.

A separate issue is that of AdS/CFT providing a fine-grained description of the bulk quantum gravity theory, that captures all the needed detail.  The possibility that 
the boundary theory only matches the bulk physics at a coarse-grained level was  raised in \refs{\FSS}, with further refinements in \refs{\GGP\GaryMI\GaGiAdS-\Isit}; ref.~\Isit\ in particular further develops the question of the detailed map needed for a true equivalence, and discusses some of the questions surrounding the possibility that AdS/CFT does not capture all details of the bulk dynamics.  

Indeed, approaching the question from a different angle, if an acceptable resolution to the black hole information problem cannot be given in AdS/CFT, for example through lack of describing the correct details of black hole evolution or through some discrepancy regarding the calculation of $\sbh$, that would provide further evidence against an exact correspondence.

There is also a microstate counting arguments that yields $\sBH$, given by Strominger and Vafa\refs{\StVa}.  However, this applies to the case of five-dimensional black holes that are both extremal, and in the weak-coupling regime; both of these characteristics separate them from the present discussion of highly non-BPS, strongly gravitationally-interacting black holes.  (Likewise considerations of, {\it e.g.},  \SenBA, apply to the extremal case, though see \SenDW.)  One would also have to interpret evidence based on duality, from comparing thermal $D0$ configurations with super-Yang Mills calculations (see {\it e.g.} \refs{\HanadaEZ}).

\subsec{Thermodynamical/statistical constraints}

Classical black hole thermodynamics has become a well-established subject, with elegant relations including that between the Bekenstein-Hawking entropy and the Hawking temperature (see \WaldLiv\ and references therein).  But, it is precisely such a semiclassical analysis that gets us into the information conflict to begin with.  
We can ask whether one could establish any contradiction resulting from $\sbh<\sBH$, like for example a violation of the second law of thermodynamics.  However, one argument against such a contradiction is that with a consistent microscopic accounting for the states of a quantum system, the quantum analog of the H-theorem and related considerations imply thermodynamic quantities behave as they should\refs{\RaMu}.

A related question, if $\sbh\neq\sBH$, is the meaning of the latter.  Classical black hole thermodynamics is remarkable, but it is also remarkable that the subject has developed as far as it has {\it without} a detailed calculation of the number of microstates of a black hole.  In seeking alternative explanations, one possibility is that $\sBH$ characterizes the properties of the near-black hole atmosphere and Hawking radiation, which are closely connected to the semiclassical geometry, and easily interact with the exterior, but that it does not count the number of internal states of a black hole.  A related viewpoint has been expressed earlier\refs{\Sorkin}, where it was observed that the second law makes no reference to conditions inside a black hole.\foot{For related discussion, see \JMR, where $\sbh>\sBH$ is advocated.}   Also, statistical mechanical systems certainly exist that, when considering one set of degrees of freedom, have a temperature, but are not in equilibrium; for example there are systems with two temperatures for different degrees of freedom\BoLa.  

Of course there is one type of microscopic calculation that {\it does} yield an entropy $\sBH$, namely the calculation of the entanglement entropy across the horizon\Sredent, if the cutoff on the theory is the Planck scale.  However, we should be suspicious of this argument, as a similar argument holds in flat space.  Indeed, in the semiclassical picture most of the modes that enter this calculation are not ``active" in the physics; the condition that determines the Hawking (or Unruh) vacuum is precisely that infalling observers see high-energy modes in the vacuum state.  Given these considerations, it is not clear how this calculation can be given an operational or measurable meaning; in counting modes that are not actually physically excited, it could be a red herring.

\subsec{Black hole pair production}

An important outstanding problem for quantum gravity is to calculate $\sbh$.  Another place where $\sBH$ is seemingly associated with the number of microstates of a black hole is in pair-production of charged black holes.  Intriguingly, there it has been found that the pair production rate of black holes, through an instanton process analogous to Schwinger production, contains a factor\refs{\GGS,\DGGH} $e^{\sBH}$.  This suggests a consistent picture where the number of produced black hole states is precisely this factor.  More specifically, the calculation where $e^{\sBH}$ enters is that of production of non-extremal black holes, connected by an Einstein-Rosen bridge. 

However, closer inspection\refs{\DGGH,\WABHIP} reveals that the calculation, which relies on a euclidean continuation of the classical near-horizon geometry, is not necessarily under control quantum-mechanically.  Ordinarily, one would expect the functional integral over fluctuations about the saddlepoint geometry to count the number of quantum states, but the route to such a consistent calculation is not clear, and the relationship between approximations of this calculation and the actual quantum states of the black hole is also not clear.  

One can argue\WABHIP\ that the calculation of this production rate contains a factor $Tr e^{-\beta H}$  for a black hole in a thermal bath, at a temperature related to the acceleration of the black hole.  Thus, using present methods, the state-counting problem is not indistinct from the problem of doing thermal calculations, which was already addressed in the preceding subsection.

\newsec{Concluding comments}

While it seems a big step to accept the statement that $\sbh<\sBH$, given the rich history of black hole thermodynamics and many appearances of $\sBH$, a broad class of models that unitarize black hole disintegration appear to imply this.  In the absence of information-transferring dynamics that matches the Hawking energy flux, a primary alternative to giving such a unitary description of black holes is, however, to accept the ``firewall" picture advocated in \AMPS.  One way or another, we are forced to give up one or more important principles.  Not only does the firewall picture apparently give up locality, in the same fashion as the scenarios considered here \refs{\BHMR\NLvC,\modelsone\modelstwo\GiSh\NVNL-\NVNLtwo}, but it requires a finely-tuned departure from locality, so information transfers just to the horizon of large black holes, but not a Planck length further.  Moreover, this picture appears to tell us that a black hole transitions into a kind of very un-blackhole-like massive remnant, which represents a violent departure from the expected semiclassical geometry near the horizon of a large black hole, particularly for infalling observers, which it apparently annihilates.  Finally, there is presently no underlying dynamical description of such firewall behavior, and moreover there is no clear way to derive black hole thermodynamics, and in particular the statement that $S_{\rm firewall}=\sBH$.  For these reasons, the scenario of non-violent nonlocality, which in the models considered here yields a super-Hawking flux and $\sbh<\sBH$, seems worthy of further exploration, even if it means giving up a statistical interpretation of $\sBH$.  Statistical considerations offer the opportunity to further refine or constrain such scenarios.

\bigskip\bigskip\centerline{{\bf Acknowledgments}}\nobreak

I thank O. Blaes, R. Bousso, J. Hartle, G. Horowitz, J. Langer, S. Shenker, M. Srednicki, and A. Wall for helpful conversations, and D. Marolf for helpful conversations and for comments on a draft of this paper.  This work  was supported in part by the Department of Energy under Contract DE-FG02-91ER40618 and by a Simons Foundation Fellowship, 229624.


\listrefs
\end